\documentclass[12pt]{article}
\usepackage{amsmath,amssymb,graphicx,mathrsfs,hyperref,slashed}
\newcommand{\be}{\begin{equation}}
\newcommand{\ee}{\end{equation}}
\newcommand{\bea}{\begin{eqnarray}}
\newcommand{\eea}{\end{eqnarray}}

\def\({\left(} \def\){\right)}

\renewcommand{\baselinestretch}{1.4}
\begin{document}
\title{\vspace{-1.8in}
{Emergent  horizon,  Hawking radiation and  chaos in the collapsed polymer model of a   black hole}}
\author{\large Ram Brustein${}^{(1)}$,  A.J.M. Medved${}^{(2,3)}$
\\
\vspace{-.5in} \hspace{-1.5in} \vbox{
 \begin{flushleft}
  $^{\textrm{\normalsize
(1)\ Department of Physics, Ben-Gurion University,
    Beer-Sheva 84105, Israel}}$
$^{\textrm{\normalsize (2)\ Department of Physics \& Electronics, Rhodes University,
  Grahamstown 6140, South Africa}}$
$^{\textrm{\normalsize (3)\ National Institute for Theoretical Physics (NITheP), Western Cape 7602,
South Africa}}$
\\ \small \hspace{1.07in}
    ramyb@bgu.ac.il,\  j.medved@ru.ac.za
\end{flushleft}
}}
\date{}
\maketitle
\begin{abstract}

We have proposed that the interior of  a macroscopic Schwarzschild black hole (BH) consists of highly excited,  long, closed, interacting strings and, as such,  can be modeled as a collapsed polymer.  It was previously shown  that the scaling relations of the collapsed-polymer model agree with those of the BH. The current paper further substantiates this proposal
with an investigation into some of its dynamical consequences. In particular, we show that the model predicts, without relying on gravitational effects,  an emergent horizon. We further show that the horizon fluctuates quantum mechanically as it should and that the strength of the fluctuations is inversely proportional to the BH entropy. It is then demonstrated that the emission of  Hawking radiation is realized microscopically by the quantum-induced escape of small pieces of string, with the rate of escape and the energy per emitted piece  both parametrically matching the Hawking temperature.  We also show, using standard methods from statistical mechanics and chaos theory, how our model accounts for some other known  properties of BHs. These include  the accepted results for the  scrambling time and the  viscosity-to-entropy ratio, which in our model apply not only at the horizon but throughout the BH interior.

\end{abstract}
\newpage
\renewcommand{\baselinestretch}{1.5}\normalsize

\section{Introduction}

The traditional narrative, based on classical general relativity, is that the interior of a  Schwarzschild black hole (BH) is a region of mostly empty space surrounding a highly dense, very small and classically singular core. It has recently been suggested that this
picture is not only flawed but spectacularly incorrect. This current narrative can be simply summarized: There is no interior to even consider. Spacetime rather ends at the BH horizon either as a matter of principle, as in the fuzzball model of BHs \cite{Mathur1,Mathur2,Mathur3,Mathur4} (also \cite{otherfuzzball}), or in practice, as a ``firewall'' of high-energy particles surrounding the horizon \cite{AMPS} (also \cite{Sunny,Braun,MP}). We, however, have been advocating a rather different point of view: The interior exists and  is not empty; it  is rather filled  with a hot bath of  long, closed, interacting strings \cite{strungout}. Such an interior cannot be described in terms of a semiclassical spacetime metric, and we  have proposed that the appropriate description of this state is in terms of a collapsed polymer.

The basis for our reasoning is that the interior of a BH must be in a highly quantum state when it is described in terms of the spacetime fields at infinity \cite{density}.
Accepting that BHs evolve in a unitary manner and that they can be in a (nearly) pure state, it follows that the interior state must be the purifier of the emitted Hawking radiation. Therefore, the non-vanishing eigenvalues of the density matrix of the emitted radiation and those of the interior state are equal.  The emitted radiation is in a highly quantum state because the emission rate of BHs is equal to the inverse of the wavelength of the emitted radiation. This is distinct from standard black bodies, such as the Sun, for which the emission rates are much larger than the inverse of the radiation wavelengths.  Therefore, the emitted radiation from BHs is  characterized by its small occupation numbers in the basis of the asymptotic fields. It follows that the state of the interior must, similarly, posses small occupation numbers.   Further, if a mean-field description of this state does exist, it must be in terms of a strongly correlated system or, equivalently, an entropically dominated
one to account for the BH's unorthodox thermodynamics; in particular, its non-extensive entropy and energy.

Once it is accepted that the BH interior does have such a highly quantum
description, one is almost inevitably led to a high-temperature theory of fundamental strings. As has been frequently discussed in the literature ({\em e.g.}, \cite{FV,MT,AW,Deo,BV,HP}), the partition function for such a theory   has an exponentially large number of closely packed states and its canonical ensemble is subject to large fluctuations that diverge at the Hagedorn temperature. In other words, this  is a theory of highly quantum matter. Moreover, the equation of state for high-temperature string theory is famously $\;p=\rho\;$, which is equivalent to $\;s=\sqrt{\rho}\;$  in suitable units. Here $p$, $\rho$ and $s$ are respectively pressure, energy density and entropy density. Thus,  $s$ is as large as it can be in comparison  to $\rho$, implying  entropic dominance.

The BH has the same equation of state, $\;s=\sqrt{\rho}\;$, if  the interior matter is distributed so as to saturate the Bekenstein--Hawking entropy--area law \cite{Bek,Hawk} for any interior, closed surface \cite{inny}. The same equation of state also signals the saturation of the causal entropy bound \cite{CEB} which implies, in turn, the breakdown of semiclassical physics \cite{maxme}. In this way, one can see how the notions of highly quantum and entropically dominant are indeed synonymous.

What we have proposed is that the BH interior should be modeled as a Schwarzschild-sized,  bound and  metastable (but classically stable) state of interacting, long, closed strings at a temperature
approaching   the Hagedorn temperature from above. Exactly at the Hagedorn temperature,  the free energy identically  vanishes and consequently the entropy density and the energy density are equal, $\;s=\rho\;$.   As just pointed out, we are considering a high-temperature phase approaching the Hagedorn temperature from above. For this phase,  Atick and Witten \cite{AW} showed that the free-energy density scales as $F/V\sim T^2$, which implies $\;\rho=E/V\sim  T^2\;$ and $\;s=S/V\sim T\;$. Hence, up to an order-unity numerical factor, $\;s=\sqrt{\rho}\;$ is indeed the relation between the entropy density and the energy density.

As discussed at length in \cite{SS,LT}, summarized in \cite{strungout} and reviewed in the next section, the resulting picture is one in which the bound state is dominated by  $\ln{N}$ long strings, where $N$ is the total length of the strings in string units (or, equivalently, the number of ``string bits"). This picture differs somewhat from the  commonly held view that a single long string is entropically favored. The latter scenario applies only when the strings are placed in a very large volume.  The difference then comes about from finite-volume effects modifying the  corresponding  equilibrium state.

Our proposal is different than the traditional BH--string correspondence \cite{corr1,corr2,corr3}, for which the duality  between BH and string thermodynamics depends on having a BH whose Schwarzschild radius $R_S$ matches the fundamental string length $l_s$. But this requirement cannot possibly  apply to macroscopic BHs. The main difference between our proposal and the BH--string correspondence is that, in our setup, there are two distinct scales of energy. One  scale is the
string-tension energy, which is given by $\;E_{ten}=N/l_s\;$ and is the microscopic energy that would be attributed to a single free string of total length $N l_s$. This energy scale is related via the first law to the Hagedorn temperature $\;T_{Hag}=1/l_s\;$ and  the string  entropy $N$.~\footnote{We mostly leave out order-unity numerical factors and fundamental constants,   often set $\;l_s=1\;$ and assume $d+1$ dimensions  with $\;d\geq 3\;$.} The other  scale is the  bound-state energy $E_{bound}$ which, together with the tension energy, accounts for the interaction energy $E_{int}$ of the strings, $\;E_{bound}=E_{ten}-E_{int}\;$. This is the energy that an external observer measures, having no direct access to the interior, and it  is the energy associated
with the Hawking temperature $\;T_{Haw}=E_{bound}/N\;$. So that, if $l_P$ is the Planck length and $\;M_{BH}=(1/l_P)\;(R_S/l_P)^{d-2}\;$ is the mass of the BH,
our proposal  requires that $\;M_{BH}=E_{bound} < E_{ten}\;$, rather than a correspondence point where $\;M_{BH}=E_{ten}\;$ as  in \cite{corr1,corr2,corr3}.

We  would now like to go beyond the  thermodynamic scaling relations of our previous work   and find out whether other known features of  BHs are shared by the collapsed-polymer model.
This is the purpose of the current paper.

The most prominent feature of a  BH is that it has a horizon. By ``horizon'', we mean  a generic term describing a classically causal barrier  to the external world. In semiclassical gravity, the horizon is described from the exterior of the BH in geometric terms, and one can make the distinction among the various geometric descriptions such as an event horizon, a particle horizon or an apparent horizon.  In our model, a  geometric description of the interior is lacking, but one can still adopt the equivalent  representation of gravity as a dynamical effect in flat space.  Flat-space coordinates, $t$, $r$, {\em etc.}, would then  represent fiducial coordinates; essentially, labels for the constituent objects (strings). From this point of view, gravity is emergent; it is an effective macroscopic description of the microscopic forces between  constituents. The geometric description (if it exists) can replace the macroscopic gravitational force.  When gravity is semiclassical, both geometric and dynamical descriptions co-exist.

Lacking a geometrical description of the spacetime, we have chosen the dynamical representation and applied it throughout the paper. What  we would then  like to understand is if (and how) a horizon emerges from the  perspective of the polymer. We find that the casual barrier can be described in the  fiducial coordinates of flat space as (approximately) a  surface of constant $\;r=R_S\;$ from which pieces of string cannot escape classically.

Further, as was argued by Bekenstein \cite{Bek} and later proved by Hawking, matter can still break through this causal barrier  but can  only do so quantum mechanically. This rasies the question of how, for our model, does  the quantum-mechanical emission of Hawking radiation come about. A crucial consistency check
for the polymer is that the emission of  Hawking particles should occur at both a  rate and energy release that matches $T_{Haw}$. We would also like to know if the emergent horizon fluctuates quantum mechanically and, if so, compare the strength of these fluctuations to that of previous works \cite{RM,RB,flucyou}.

Another distinctive feature of BHs is that they are  maximally fast scramblers of information \cite{HayPres,Suss,Suss2}. This leads to another important litmus test for our model: It must reproduce the minimally allowed scrambling time. This is basically the same as realizing the maximum possible value of the Lyapunov exponent, which also makes BHs the most chaotic of all systems \cite{MaldShenk}. Closely related to the idea of a BH being an ideal scrambler and maximally chaotic  is that its viscosity-to-entropy ratio is minimized \cite{membrane,PSS,KSS}, thus providing another essential check.

After a review of our proposed model, the remainder of  the paper puts it to the test by investigating the questions posed above. We find that it passes all of them with ``flying colors'', providing a direct and simple physical explanation in all cases.

\section{Review of the polymer model}

Much of the content of this section has been already  covered in \cite{strungout}. We include it here for completeness.

\subsection{Highly excited strings in a bounded region}

We wish to describe a bound state of long closed strings. The idea is that the closed strings occupy a bounded region in space whose size is determined dynamically. A more figurative description of the proposed bound state is that of a ``quantum star" made of fundamental strings in the Hagedorn phase or simply a ``string ball".

We start our discussion by following the descriptions of  Salomonson and Skagerstam (SS) \cite{SS}, and  Lowe and Thorlacius (LT) \cite{LT} of highly excited strings in a bounded region. Their description relies on two properties of interacting, long strings: (1) That the total length is conserved due to the conservation of energy and (2) that the probability for interaction is proportional to the length of the string. In the particular setting of the SS--LT model ---  where
interactions are described by the splitting and joining of strings ---
 only the three-string interactions are important, while  the four-string  and higher-order interactions can be neglected. The SS--LT model was later validated by  detailed investigations using the ``thermal-scalar" formalism \cite{HP}, the world-sheet description of (open) bosonic strings \cite{DV} and of closed-strings \cite{chialva}.

The  LT description is particularly useful for  motivating our proposal for the BH interior. These authors considered a high-temperature collection of interacting, closed strings in a finite volume such that the total length of string is much longer than the spatial dimensions of the enclosing volume. They consider the volume of the confining region or  ``box'' to be fixed; for example, strings moving on a toroidal space of a fixed size. We regard  the  finite volume of the bound state as being  due to  the (attractive) interactions of  the strings. So that, in our case, the interaction energy  is parametrically the same as the string tension. As discussed later, it is also true for our case that the three-string interactions are important, and the four-string and higher-order interactions can be neglected.

\begin{figure*}[!ht]
\begin{center}
\includegraphics[width=.45\textwidth]{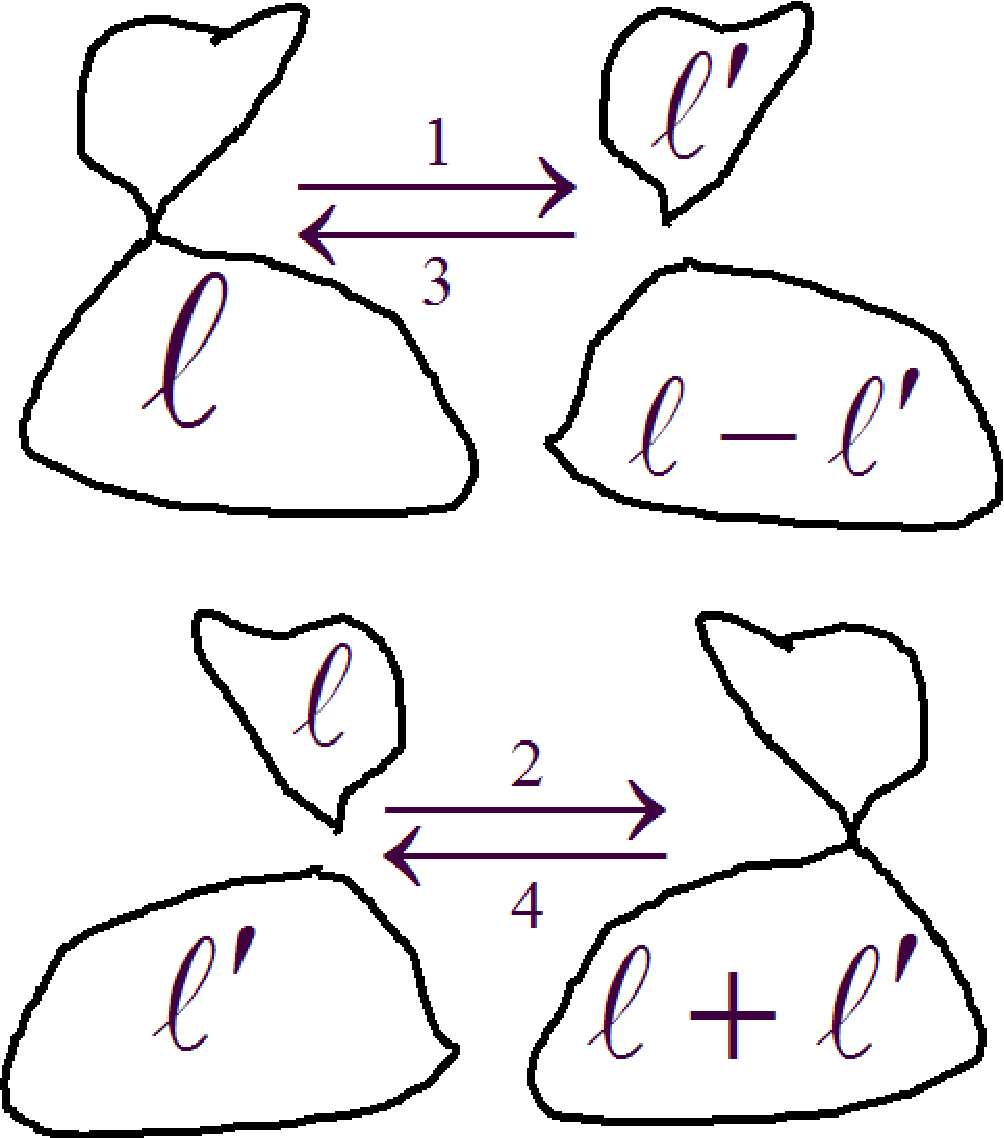}
\caption{Strings joining and splitting. The number 1 refers to
the first term on the right in Eq.~(\ref{Boltz2}), and so on.}\label{FigLT}
\end{center}
\end{figure*}

Let us briefly sketch  the pertinent results of  LT, who considered the Boltzmann equation for long, closed strings when only three-string interactions are relevant. In  $\;l_s=1\;$ units, this is
\cite{LT}
\bea
 \frac{dn(\ell)}{d t} \;=\; k \frac{g^2}{V}\Big[&-&\frac{1}{2}\ell^2 n(\ell)-\int^{\infty}_0 d\ell'\;\ell'n(\ell')
\ell n(\ell)
 \nonumber \\
&+&\frac{1}{2}\int^{\ell}_{0} d\ell'\;\ell'(\ell-\ell')n(\ell')n(\ell-\ell')
+ \int^{\infty}_{\ell} d\ell'\; \ell' n(\ell')\;\Big]\;,
\label{Boltz2}
\eea
where $n(\ell)$ is the average number of strings of length $\ell$,   $k$ is a numerical factor and each of  the four terms on the right represents a different type of three-string interaction. See Fig.~\ref{FigLT}. The first term  represents the possibility that a string of length $\ell$ splits into two shorter ones, while the second term accounts for  two strings of lengths $\ell$ and $\ell'$ joining together to form one of length $\ell+\ell'$. The third term results from two shorter strings joining to make a longer string of length $\ell$ and, finally,  the last term accounts for a longer loop splitting into two shorter ones, one of which is of length $\ell$. Higher-order interactions are suppressed by the smallness of the string coupling and/or the largeness of the volume.

The equilibrium condition $\;{ dn}(\ell)/dt=0\;$  leads to the distribution
$
n(\ell) \;=\; \frac{1}{\ell} e^{-{\ell}/{L}}\;,
\label{above}
$
where $\;L=\int_0^{\infty}d\ell\;\ell n(\ell)\;$.
The strings are distributed such that, on average, there is one string at each length interval from
$\;\ell=\ell_i\;$  to $\;\ell=e \ell_i\;$ (starting with $\;\ell_i=l_s\;$ and
ending with $\;\ell_i\sim L/e$) up to $\;\ell=L\;$.

Let us define $N$ as the number of ``string bits" of length $l_s$, so that $\;N=L/l_s\;$.  The string entropy is then essentially that  of an $N$-step random walk, which is the logarithm of the number of different walks. If, for example, each step is $\pm 1$, then $\;S=\ln{2^N}\sim N\;$. The above distribution
implies that most of the bits will be found on the longest strings, which  will then make the dominant contribution to the entropy and energy,  both of which  scale with the total length of string.

\subsection{A bound state of highly excited strings: The collapsed polymer}

Our next objective is to describe the BH as a bound state of highly excited strings. Here, strong interactions lead to the formation of the bound state, whose volume is determined dynamically  rather than by an external ``box".  As in the LT--SS description and as mentioned in the Introduction, flat-space coordinates, $t$, $r$, {\em etc.}, will represent fiducial coordinates; labels for the strings in spacetime. In particular, $r$ is the radial fiducial coordinate with
$\;r=0\;$ located at
the center of the state. We will consider spherically symmetric configurations and   let  $\;r=R_S\;$ denote the finite spatial extent of the state.  Although this need not be the Schwarzschild radius {\em a priori}, we will later show that it is parametrically equal to the Schwarzschild radius of the corresponding BH.

\begin{figure*}[!ht]
\begin{center}
\includegraphics[width=.45\textwidth]{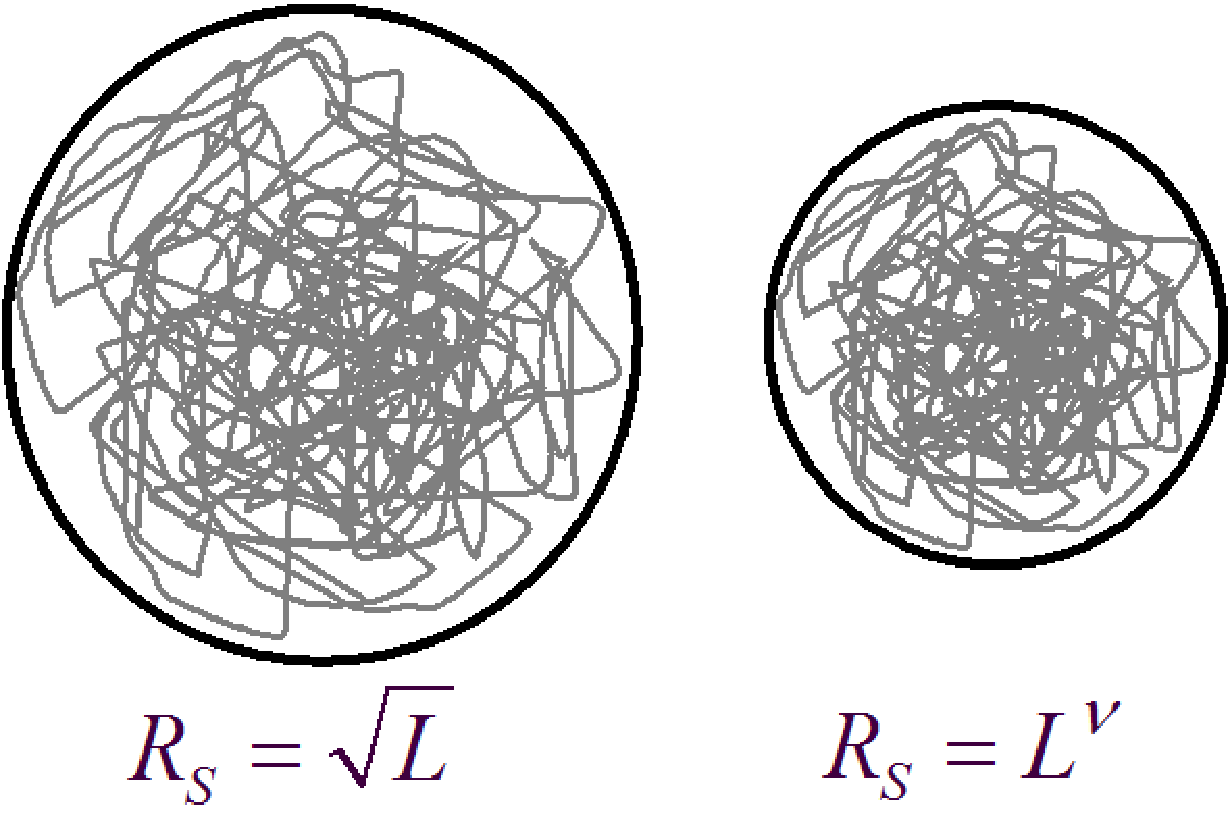}
\caption{Strings in a bounded region. Free strings (left) occupy a larger region than interacting strings (right).}\label{collapsedpolymer}
\end{center}
\end{figure*}

When the strings are free and put in a large-enough volume, the radial size $R_S$ of the region  is determined by the random-walk scale, which is
 $\;R_S\sim \sqrt{N}\;$ (in $\;l_s=1\;$ units)  for a   string of total length   $N$. The joining and splitting of the closed strings lead, effectively, to an attractive interaction as they tend to make the strings occupy a smaller region in space. These attractive interactions therefore lead to a smaller value of $R_S$  \cite{HP,DV}. The free and interacting cases are depicted in Fig.~\ref{collapsedpolymer}. In general, in the absence of another scale besides that of $l_s$ (which is set to unity),  we expect that $\;R_S\sim N^\nu\;$ for some $\nu$.
The value of $\nu$   is determined by the strength of the string interactions. Since $\;N\sim R_S^{1/\nu}\;$, the entropy will not, in general,  be extensive in the volume. An area law, as in the case of BHs,  implies that  $\;\nu=1/(d-1)\;$.

Such a scaling relation between the entropy and the spatial size of the state appears in the Flory--Huggins theory of polymers \cite{flory}. (See also the books by De Gennes \cite{bookdg} and Doi and Edwards \cite{bookde}.) This theory is reconsidered in \cite{degennes} and reviewed, for example, in \cite{polymer}. The transition temperature when the polymers become tensionless is known as the Flory temperature and $\nu$ is called the Flory exponent. The linear size $R_S$ is known as the gyration radius of the polymer. In  the polymer literature, the case of attractive interactions  is  referred to as ``negative  excluded volume". In this case, the gyration radius of the polymer is smaller than $\sqrt{N}$ and the resulting state  is  called a ``collapsed polymer".

The theory of collapsed polymers therefore provides  an effective description of the bound state of highly excited strings, along with the appropriate adaptations
which we now explain. The Flory effective free energy for the case of a collapsed polymer (see, {\em e.g.}, \cite{polymer})  is given by
\be
-\left(\frac{F}{VT}\right)_{\rm CP}\;=\;c - \frac{1}{2} \upsilon c^2  +{\cal O}[c^3]\;.
\label{cp}
\ee
Here, $c$  is the entropy density (monomer concentration)  and $\upsilon$ is the absolute value of the negative excluded volume. The relevance of the higher-order terms is controlled by the value of $\upsilon$. If the negative excluded volume is parametrically smaller than the volume $\;V\sim R_S^d\;$  occupied by the polymer, then the higher-order interaction terms are not important, just as in the
SS--LT model.

The Flory theory then  suggests an effective free energy for highly excited strings at a temperature close to but above the Hagedorn temperature, $\;T\gtrsim T_{Hag}$.  In terms of the ``bit concentration"
or entropy density $\;c=N/V\;$ of the strings, this is
\be
-\left(\frac{F}{V T_{Hag}}\right)_{strings}\;= \;\epsilon c - \frac{1}{2} \upsilon c^2 +{\cal O}[c^3]\;
\label{FES}
\ee
or, equivalently,
\be
-\left(\frac{F}{T_{Hag}}\right)_{strings}\;=\; \epsilon N - \frac{1}{2} \frac{\upsilon}{V} N^2 +{\cal O}\left[\frac{N^3}{V^2}\right]\;.
\label{FES1}
\ee
We will treat the Hagedorn temperature $T_{Hag}$  as equal to the string energy (ignoring an order-one numerical factor), so that  $\;T_{Hag}= 1\;$.

We now would like to discuss the  excluded volume $\upsilon$ and
the additional (dimensionless) parameter $\epsilon$ for
the specific case of  highly excited strings. Let us begin with the latter
parameter.

For this case of excited strings, we know that the leading-order expressions for the energy and entropy
both scale with the number of string bits $N$; that is,
$\;S\sim N\;$ and $\;E\sim N\;$ in string  units.
This is because the leading-order values  should be
comparable to those of the free-string case.
Next consider that the free energy  $\;F=-(ST-E)\;$ is, at leading order,  expected to vanish at the Hagedorn temperature.   It then follows that, for $T$ close to $T_{Hag}$,
the difference $ST-E$  should be  a number which is  parametrically smaller
than $N$.  We have done, above, is to introduce a dimensionless parameter $\epsilon$ that  parametrizes  this small  number.  In fact, in light of Eq.~(\ref{FES1}),
one can deduce that this parameter must of the form  $\;\epsilon=(T-T_{Hag})
/T_{Hag}\ll 1\;$.  This follows from  $T \gtrsim T_{Hag}= 1\;$ and then
$\;\epsilon N\sim -F/T_{Hag}= (ST-E)/T_{Hag}\sim (NT-NT_{Hag})/T_{Hag} =
N(T-T_{Hag})/T_{Hag}\;$.

The second term on the right of  Eq.~(\ref{FES1}) is meant to account
for  the effect of the  interactions, which
is parametrized  by the excluded volume $\upsilon$ in the collapsed-polymer framework.
In our case, the strength of the interactions
is determined by $g^2$, and  any one  interaction is
due to either  the  joining of a pair of  strings  or the  splitting
 of a  single string. These actions can  take place
 whenever two bits of string  occupy the same ``point'' in spacetime.
Since the spacetime in question can be viewed as a lattice with a spacing
of size  $l_s$,
the excluded volume can be  identified  as
$\;\upsilon = g^2 l_s^d=g^2\;$, and  the density of
coincident bits is then given by  $g^2 N^2/V$.

The effective free energy
of the excited state of strings  can now be expressed as
\be
-\left(\frac{F}{V T_{Hag}}\right)_{strings}\;= \;\epsilon c - \frac{1}{2} g^2  c^2 +{\cal O}[c^3]\;
\label{FES2}
\ee
or
\be
-\left(\frac{F}{ T_{Hag}}\right)_{strings}\;= \;\epsilon N - \frac{1}{2} g^2  \frac{N^2}{V} +{\cal O}\left[
\frac{N^3}{V^2}\right]\;,
\label{FES3}
\ee
where, as discussed above,  $\epsilon$  can  be identified
as the  deviation of the temperature from the Hagedorn temperature, $\;\epsilon=(T-T_{Hag})/T_{Hag}=T-1\;$. The value of $\epsilon$ need not be conjectured; its
equilibrium value follows immediately from the minimization of the free energy,
which gives (see below) $\;\epsilon=g^2 N/V\;$.
Then, insofar as
the string state forms  into a BH, it is  clear via  the BH area law
$\;N\sim R_S^{d-1}/g^2\;$ that,
in string units,  $\;\epsilon\sim 1/R_S\sim T_{Haw}\;$.

The basic form $\;-F=ST-E\;$ tells that $\;\epsilon\sim T_{Haw}\;$ can
indeed be viewed as an effective temperature for the bound state.
Alternatively, $\epsilon$ can also be viewed as the relative increase in the excluded volume. That is, we can divide Eq.~(\ref{FES1}) through by $\epsilon$
and absorb  $1/\epsilon$ into $\upsilon$. From this point of view, $1/\epsilon$
tells us how much that the excluded volume increases as the temperature
increases  from the   Hagedorn value.

From the free energy   in either  Eq.~(\ref{FES2}) or~(\ref{FES3}), one can obtain the equilibrium condition
which follows from $\;\frac{\partial F}{\partial c}=0\;$. The solution, $\;c=\epsilon/g^2\;$, is valid only for finite coupling and is therefore ``non-perturbative". In terms of $N$, the same equilibrium condition is
\be
\label{consol}
\frac{N}{V} \;=\; \frac{\epsilon}{g^2}\;.
\ee

One can now evaluate the  entropy density, energy density and pressure via the standard thermodynamic definitions. The results are as follows:
\bea
s &=& -\left.\frac{1}{V}\frac{\partial F}{\partial \epsilon}\right|_{c=\epsilon/g^2}\;=\; \frac{\epsilon}{g^2}\;,
\nonumber \\
\rho &=& \left[\frac{F}{V} + \epsilon s\right]_{c=\epsilon/g^2} \;=\; \frac{1}{2}\frac{\epsilon^2}{g^2}\;,
\nonumber \\
p&=& -\left.\frac{\partial F}{\partial V}\right|_{c=\epsilon/g^2} \;=\; \frac{1}{2}\frac{\epsilon^2}{g^2}\;.
\eea
Here, we have obtained  the signature equation of state  for a
of  high-temperature bath of closed  strings, $\;p=\rho\;$.

In addition, one can calculate the effective tension and confirm that it vanishes at (and near) the Hagedorn or Flory temperature as expected,
\be
\sigma\;=\; \left.\frac{\partial F}{\partial L}\right|_{c=\epsilon/g^2} \;=\;\left.\frac{\partial F}{\partial N}\right|_{c=\epsilon/g^2} \;=\;0\;.
\ee

We can  independently confirm the scaling relation $\;\epsilon\sim 1/R_S\;$ from the fact that, at long distances, the string interactions are dominated by gravity \cite{HP,DV}. Hence, the total interaction energy of the bound state of strings is parametrically equal  to its total gravitational energy,
\be
g^2 \frac{N^2}{V} \;\sim\; G_N \frac{E^2}{R^{d-2}_S}\;,
\label{grav-energy}
\ee
where the left-hand side follows from the interaction  term in the free energy~(\ref{FES3}), or, equivalently, from an inspection of the LT Boltzmann Equation~(\ref{Boltz2}).
The right-hand side is the Newtonian potential with
$\;G_N=l_P^{d-1}=g^2 l_s^{d-1} = g^2\;$.
The energy $E$ on the right side of Eq.~(\ref{grav-energy}) must be the bound-state energy $\;E=\rho V\sim \epsilon N\;$, and so
\be
\frac{R^{d-2}_S}{\epsilon^2 V}\;\sim\; 1\;.
\ee
Then, since $\;V\sim R_S^d\;$, it follows that  $\;\epsilon\sim 1/R_S\;$,
same as before.

The self-consistency of this framework necessitates
the following set of hierarchies:
$\;\epsilon\ll g^2 \ll 1\;$ and $\;g^2 N = V\epsilon \gg 1\;$  so that the BH is large in string units. These conditions ensure that the concentration
is small,
$\;c=\epsilon/g^2 \ll 1\;$, which
 validates the expansion of the free energy and means that the higher-order terms can be neglected. Additionally, these conditions  guarantee that the higher-order string interactions coming from  $\alpha^\prime$ corrections, loop corrections or their combination are indeed suppressed.  Nevertheless, they also ensure that $g^2$  is finite.

The relative strength of  $(n+2)$-string interactions is proportional to $g^{2n}$ and to a combinatoric  enhancement factor $N^n$ and suppressed by a  volume suppression factor $V^{-n}$.
In string units, the multi-string  interaction strength then goes as
\be
\lambda_{n+2}\;\sim\;  \left(\frac{g^{2}N}{V}\right)^n\;=\; \epsilon^{n}=\left(\frac{l_s}{R_S}\right)^n\;,
\ee
where we have used Eq.~(\ref{consol}). One can see that these are $\alpha^\prime$ corrections,
being proportional to powers of $l_s/R_S$.  Similarly, higher-order string-loop corrections and combinations thereof are suppressed, provided that
$\;g^2\ll 1\;$.

Let us also recall that the size of a free string of a fixed length $N$ in string units  is $\;\sqrt{N}=1/g\epsilon^{(d-1)/2}\;$, which follows from a random-walk scaling. Hence,   the bound-state size of $\;1/\epsilon\;$ is smaller than the free-string size by a factor of $g\epsilon^{(d-3)/2}\ll 1\;$; meaning that the interactions are  important, even though the higher-order interaction
terms can be safely  neglected.

The previous thermodynamic scaling relations agree with those of a BH, provided that the bound-state energy is  identified with the BH mass and the effective
temperature with the Hawking temperature. For completeness, the dictionary
goes as follows
(in units in which $\;l_s=1\;$ and using only the  dimensionless parameters $g^2$ and $\epsilon$): The volume of the interior  is given by
\be
V\;=\;\frac{1}{\epsilon^{d}}\;,
\label{bhv}
\ee
the entropy by
\be
\label{bhs}
S_{BH}\;=\;N\;=\; V\frac{\epsilon}{g^2}\;,
\ee
the tension energy by
\be
\label{bhten}
E_{ten}\;=\;N\;=\; V\frac{\epsilon}{g^2}\;
\ee
and, correspondingly,
\be
\label{bhthag}
T_{Hag}\;=\;1\;.
\ee
The total (bound-state) energy is
\be
\label{bhm}
M_{BH}\;=\;E_{bound}\;=\;V\frac{\epsilon^2}{g^2}\;
\ee
and, correspondingly, the Hawking temperature is
\be
\label{bhthaw}
T_{Haw}\;= \;\frac{E_{bound}}{N}\;=\;\epsilon\;.
\ee
In spite of the implicit presence of Planck's constant in its numerator, the Hawking temperature depends only on $\epsilon$ and not on $g^2$. This is so because Newton's constant cancels out of the ratio $M_{BH}/S_{BH}$.

\section{Emerging horizon and Hawking radiation}

As a gravitational concept, a  horizon is a geometric realization of  the fact that, classically, matter cannot escape from the interior of a  BH. That is, a horizon can be defined, classically, as a closed surface through which matter is permitted to enter but not exit. Quantum mechanically, we expect two related effects to modify this picture.

For concreteness, we will consider an apparent BH horizon. According to its standard geometric  and classical definition, an apparent  horizon is a null spacetime surface where the effects of gravity are so strong that the escape velocity equals that of the speed of light. Consequently, incoming photons will end up  orbiting this  surface, whereas massive particles would inevitably fall in. Additionally,  matter that is  already inside this surface  will never be able to escape, at least  by classical means. However, from our viewpoint, the  gravitational interactions in a curved spacetime  have been   replaced by  inertial forces  in a flat space, and so a dynamical  description of the  horizon is required.

Let us emphasize that the location of an apparent BH horizon is a physical observable since it is a measurable quantity (in principle). In \cite{Visser}, it is explained how an observer in a finite-sized laboratory can detect the horizon  by measuring  nearby tidal effects. For instance, an observer can measure the convergence of falling objects as a function of radial distance and use these measurements to determine where the curvature invariants coincide with their horizon values. This conclusion evades any violation of  Einstein's equivalence principle because the principle is   only relevant   to ultra-local measurements.

How would a bound state of highly excited strings  in a finite region of space -- a ``ball of string'' --- form a dynamical horizon?  Classically, the strings cannot split, as the spliting of strings is a quantum effect in string theory. If the strings cannot split,  bits of  string cannot break off and escape from the  ball of string.   Our new  definition of the horizon is then the smallest  spherical shell that entirely encloses the ball of string. The size of this shell can, classically,  be expected to be set by
the random-walk scale of a free string. Then, just like the standard
geometric  definition of a horizon, this dynamically formed shell effectively describes  a one-way  barrier to the external world.

Once quantum mechanics is taken into account, the standard definition of a  BH horizon becomes less sharp. The reason for this is two-fold: First, the position of the horizon is expected to fluctuate, as does, generically, any other physical observable. Second, the process of Hawking radiation.

Regarding fluctuations,  we have previously shown \cite{RM,RB,flucyou} that the quantum variance of $R_S$ scales as
\be
\label{deltar}
\frac{(\Delta R_S)^2}{\langle R_S\rangle^2} \;\simeq\; \frac{1}{S_{BH}}\;.
\ee
This  is a general result that is applicable in any dimension and
follows from simple quantum-mechanical considerations:
Whatever  quantum  wavefunction correctly describes a BH, it must yield
certain expectation values in the classical limit according to the
Bohr correspondence principle. Clearly, the expectation value of
the  ``entropy operator" ${\widehat S}$ (See \cite{RM,RB,flucyou} for the precise definition)  must correspond to  the
Bekenstein--Hawking entropy $S_{BH}$. As for the variance
of the entropy operator, standard thermodynamics dictates  that
this is given by  the  magnitude of
 the heat capacity, which is also of the order of  $S_{BH}$.
It then  follows that
\be
\frac{(\Delta {\widehat S})^2} {\langle {\widehat S}\rangle^2} \;\simeq\; \frac{1}{S_{BH}}\;,
\ee
from which Eq.~(\ref{deltar}) can be obtained  by way of  the chain rule.

That $1/S_{BH}$ acts  as a dimensionless $\hbar$ is also
apparent  in the context of the AdS/CFT duality: The quantity that controls the strength of quantum fluctuations on the field-theory side is $1/{\cal N}^2$, which on the gravity side corresponds to $1/S_{BH}$ (here,  ${\cal N}$ is the number of colors).

The second quantum effect to consider is the emission of  Hawking radiation. Quantum mechanically, matter can escape from the BH due to the Hawking process. The rate and average energy of the escaped particles is determined by the size of the BH,  $R_S$.

The purpose of this section is to  verify   that  the general picture that we just presented can be reproduced
from the specific microscopic model of  the collapsed polymer. This means establishing the emergence of a classical horizon that fluctuates quantum mechanically at the correct strength as  in  Eq.~(\ref{deltar}).
Also, that strings escape from the bound state via a quantum-mechanical effect whose rate and energy of emission agree with those of the Hawking process. Since the microscopic model does not contain gravity explicitly, this constitutes a highly non-trivial check on the viability of our proposal for the interior.

\subsection{Emerging Horizon}

Let us recall the  Boltzmann-like equation for interacting, closed
strings, Eq.~(\ref{Boltz2}). The string interaction strength is proportional to $g^2$ \cite{SS,LT}. So that, in the classical approximation, the coupling $g^2$ vanishes and pieces  of the polymer cannot  detach from its main body.
This is the simplest expression of a horizon, as the escape velocity of a piece  must be infinitely large as long as it is still connected to the polymer. The only question in this case is how far away can a piece get? Perhaps
it can still escape to infinity for all practical purposes.

A related question is to what extent is the size of the polymer  well defined?
This leads to a more precise criterion for the existence of a horizon:
The extent to which pieces of the polymer can protrude away from the main body in relation to its diameter $2R_S$.  What we then need to estimate is the strength of the  fluctuations in the position of the horizon and check that these are
  sufficiently weak.

To this end, let us recall Eq.~(\ref{FES2}) for  the
effective free energy $F$ in terms of  the polymer concentration $c$,
where $c=N/V\;$ at equilibrium.
Rearranging slightly (and using $\;T_{Hag=}=1/l_s=1$), we have
 \be
\frac{F}{\epsilon}\;=\; -cV + \frac{1}{2}V \frac{g^2}{\epsilon}
c^2 \;+\; {\cal O}[c^3]\;.
\label{poly2}
\ee

Since $c$ is conjugate to the volume $V$ in Eq.~(\ref{poly2}),
one can apply  standard statistical mechanics and determine the variance of the volume fluctuations  by twice varying
the free energy  with respect to the polymer concentration. That is,
\be
(\Delta V)^2 \;=\; \frac{\partial^2 (F/\epsilon)}{\partial c^2}
 \;=\; \frac {V g^2}{\epsilon} \;.
\ee

Using the equilibrium relation in Eq.~(\ref{consol}), $\;\epsilon=g^2 N/V\;$, and dividing
through by $V^2$,  we  now  obtain
 \be
\frac{(\Delta V)^2}{V^2} \;=\; \frac{1}{N}=\frac{1}{S_{BH}}\;.
\ee

Then,  by way of the  chain rule, the boundary or horizon fluctuations
scale according to
\be
\frac{(\Delta R_S)^2}{R_S^2} \;\simeq\; \frac{1}{N}\;,
\ee
just as expected for a BH with an entropy of $N$. Clearly, this is the smallest possible variance, being that
it  is proportional to $1/S_{BH}$ , the dimensionless $\hbar$ for this framework.
Thus, the position of the horizon is as well defined as it can be.

\subsection{Hawking radiation}

\begin{figure*}[!ht]
\begin{center}
\includegraphics[width=.45\textwidth]{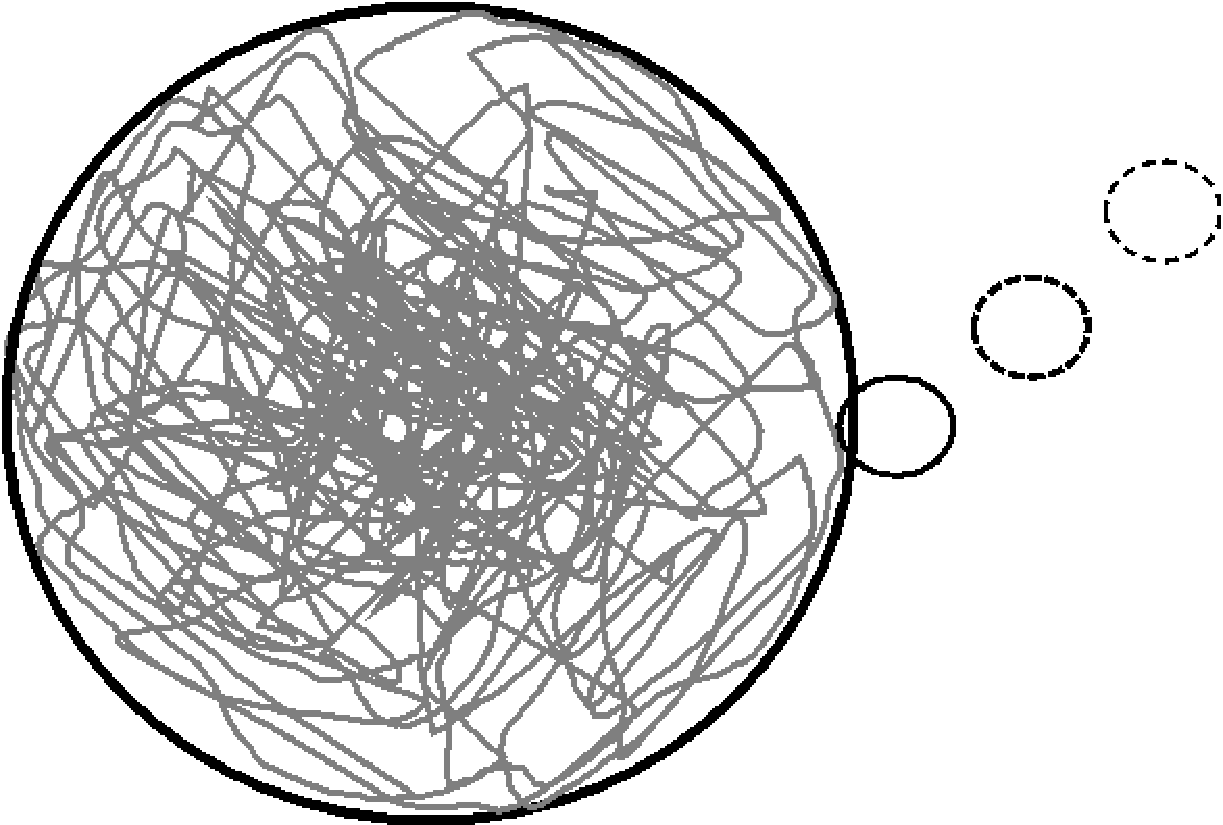}
\caption{A dynamical Hawking process: Small loops can tear off and escape from the bulk of the polymer.}\label{HawkingRadiation}
\end{center}
\end{figure*}

The viability of quantum escape can be determined
in  two steps: By,  first, calculating  the rate at which small strings escape
(see Fig.~\ref{HawkingRadiation})  and, second,  determining the  suppression factor for the rate of escape of longer strings with respect to the small strings. Throughout  this subsection, we use units in which $\;l_s=1\;$.

To address the first step, let us start by recalling  that the equilibrium solution of Eq.~(\ref{Boltz2})
for the string-number function is \cite{LT}
\be
\label{nell}
n(\ell) \;= \;\ell^{-1}e^{-\frac{\ell}{L}}\;,
\ee
where $L$ is the expectation value of the total length of string ({\em i.e.}, $\;L=N\;$
in string units). It follows that, on average, the number of the shortest strings -- loops of unit length  -- is approximately
$\;n(1)\simeq 1\;$. We can therefore assume that there is only a single short string at any given time.

The probability that a short string  can  survive  its journey  from an arbitrary location in the polymer to the outer surface depends  mainly on its mean-free-path length $l_{mfp}$. Once a small string does safely reach the surface,
it has an order-unity probability of escaping if its velocity contains an outward component. On the other hand,
 if the short string instead interacts, it will join onto a longer string, and whether or not this longer loop can
make it to the surface is a case which will be discussed separately.

It is easier to begin with the mean-free-path time,
$\;t_{mfp}=l_{mfp}/v_{CoM}\;$ with   $v_{CoM}$ denoting  the speed of the  string's center of mass.
The mean-free-path time for a string of length $\ell$ can be determined by way of
the following relation:
\be
\label{dEdt}
\frac{dE}{dt}\;=\;\frac{g^2N}{V}E(\ell)\;,
\ee
where the left-hand side is the power absorbed
by the  string because of  interactions and  $\;E(\ell)\propto\ell\;$ is its
energy. An  expression of this basic form  follows
from  $g^2N/V$ measuring the overall strength of the interactions and from
the interaction rate of any given  string being  proportional to its length \cite{SS,LT}.

We can estimate $t_{mfp}$  for a short string by defining it  as the time when
the absorbed energy  becomes of the same order as its original energy $E(1)$.
This will  happen  when
\be
t_{mfp}\;\simeq\; \frac{V}{Ng^2} \;=\;\frac{1}{\epsilon}\;,
\ee
where we have recalled the equilibrium condition $\;V/Ng^2=1/\epsilon\;$ in Eq.~(\ref{consol}).
Since a string  with a length  of order one  can be regarded as relativistic, it follows that
\be
\;l_{mfp}\;=\;t_{mfp}\;\simeq\;\frac{1}{\epsilon}\;
\label{lmfp1}
\ee
for the shortest strings.

As $\;1/\epsilon=R_S\;$, the result in Eq.~(\ref{lmfp1}) means that a  short string has an order-one probability
of making to the surface. And, since  there is (on average) only one such string at any given time,
the rate of escape of the short strings    is just the inverse of the mean-free-path time,
$\;1/t_{mfp}=\epsilon\;$. This is exactly the rate at which BHs emit their radiation because $\;T_{Haw}=\epsilon\;$ determines the emission rate of the  Hawking particles.
And so the rate of escape of  strings whose length is of order one  is
\be
\frac{1}{t_{mfp}}\;\simeq\; T_{Haw}\;.
\ee

We can therefore conclude that the emission of Hawking radiation is a consequence of
the smallest possible loops of string (on the order of one string length) escaping  from
the collapsed  polymer. This fits well with energetic considerations, as  a single bit of string will carry  a net energy of about
\be
E_{emitted}\;=\;\frac{M_{BH}}{N} \;=\;\epsilon \;=\; T_{Haw}\;
\ee
away from the bound state. Therefore, if only short strings contribute, the rate of energy emission is
\be
\label{totrate}
\frac{dE_{emitted}}{dt}\;\simeq\;\epsilon^2 \;=\; T_{Haw}^2\;.
\ee

This energy  budgeting  may seem contradictory to a just-escaped  string having  a tension of order one. However, like any particle in a bound state  or a potential well, it expends some energy in climbing out of the well. Here, we know the average of how much  energy is typically retained by such a string; it is simply $\;E_{bound}/N=\epsilon\;$.  Also note that there is effectively no Tolman redshift in this calculation because of our choice to  describe  gravity as an inertial force in a flat spacetime.

We do not mean, of course,  that an  escaped string will directly transform into  a physical Hawking particle. Once in the exterior, the string  is quickly subsumed  into the external string condensate which describes the classical gravitational field. The condensate will then rearrange to accommodate this new excitation; an action which is manifested in the physical spacetime as the emission  of a Hawking particle.  An external observer would naturally attribute this process to one of gravitationally induced pair production \cite{info}. Meanwhile, the internal polymer rearranges to accommodate the loss of the string, finally settling  into a new equilibrium state with a   slightly lower  energy. The external observer would attribute this reduction in energy to the absorption of the negative-energy pair partner of the emitted  particle.

What still needs to be checked is the fate of  strings that are parametrically longer  than the fundamental string length. If these strings escape at a significant rate, then energy will be leaking faster from the polymer than it would be for a BH. Fortunately, there is a simple statistical argument which ensures that the escape rate of long strings is exponentially suppressed.~\footnote{We thank Masanori Hanada for suggesting this argument.}

We can use entropic considerations to estimate the probability that a long string of length $\;\ell \gg 1\;$
detaches from the body of the polymer. A standard argument in statistical mechanics maintains that, near equilibrium,
the  relative probability of a string of  total length $L$ splitting into two disconnected strings ---  one of length $L-\ell$ and the other of length $\ell$ ---  decreases exponentially with the entropy of the detached string, which is equal to its length $\ell$ in string units.  It follows that the  probability of detachment goes as
\be
\frac{ P(L-\ell,\; \ell)}{P(L)}\;\sim\; e^{-\ell}\;.
\ee

It can be concluded that the total emitted energy  from all long strings of any length  is suppressed by an
exponentially small factor  with respect to that of the shortest strings.~\footnote{There will  also be an additional power-law suppression due to longer strings being fewer in number ({\em cf}, Eq.~(\ref{nell})),  having a shorter mean-free-path length
and a non-relativistic velocity.} Therefore, the long strings make a negligible
contribution.

\section{Chaos, fast scrambling  and viscosity}

The main objective of  this section is  to understand how (and if) the collapsed polymer
is able to reproduce the maximally fast scrambling rate  \cite{HayPres,Suss,Suss2} and
the minimally allowed viscosity-to-entropy ratio \cite{membrane,PSS,KSS} of a BH. From the BH perspective, these properties pertain  to just  the horizon, but we will be able to make a novel prediction about the latter one in the interior. The key idea in this section is the use of  standard statistical mechanics to relate the polymer  free energy to the various quantities of interest.

\subsection{Chaos and fast scrambling}

In the context of BHs, the scrambling time is meant to be the typical  time scale for  a  disturbance that is induced by
in-falling  matter to  disperse over the entire horizon surface. Underlying this definition is the premise that an external observer can  never  see matter or energy fall into the BH --- so that, as far as she is concerned, any such disturbance is forever  trapped on the horizon.~\footnote{It is more conventional to frame the scrambling time in terms of  the ``stretched horizon'', which is taken to be a few Planck lengths outward  from the actual horizon.} But the  internal  perspective, which must be adopted in our framework,  is necessarily somewhat different and more similar to the way that the scrambling time is defined in systems which can be accessed by an observer.

\begin{figure*}[!ht]
\begin{center}
\includegraphics[width=.45\textwidth]{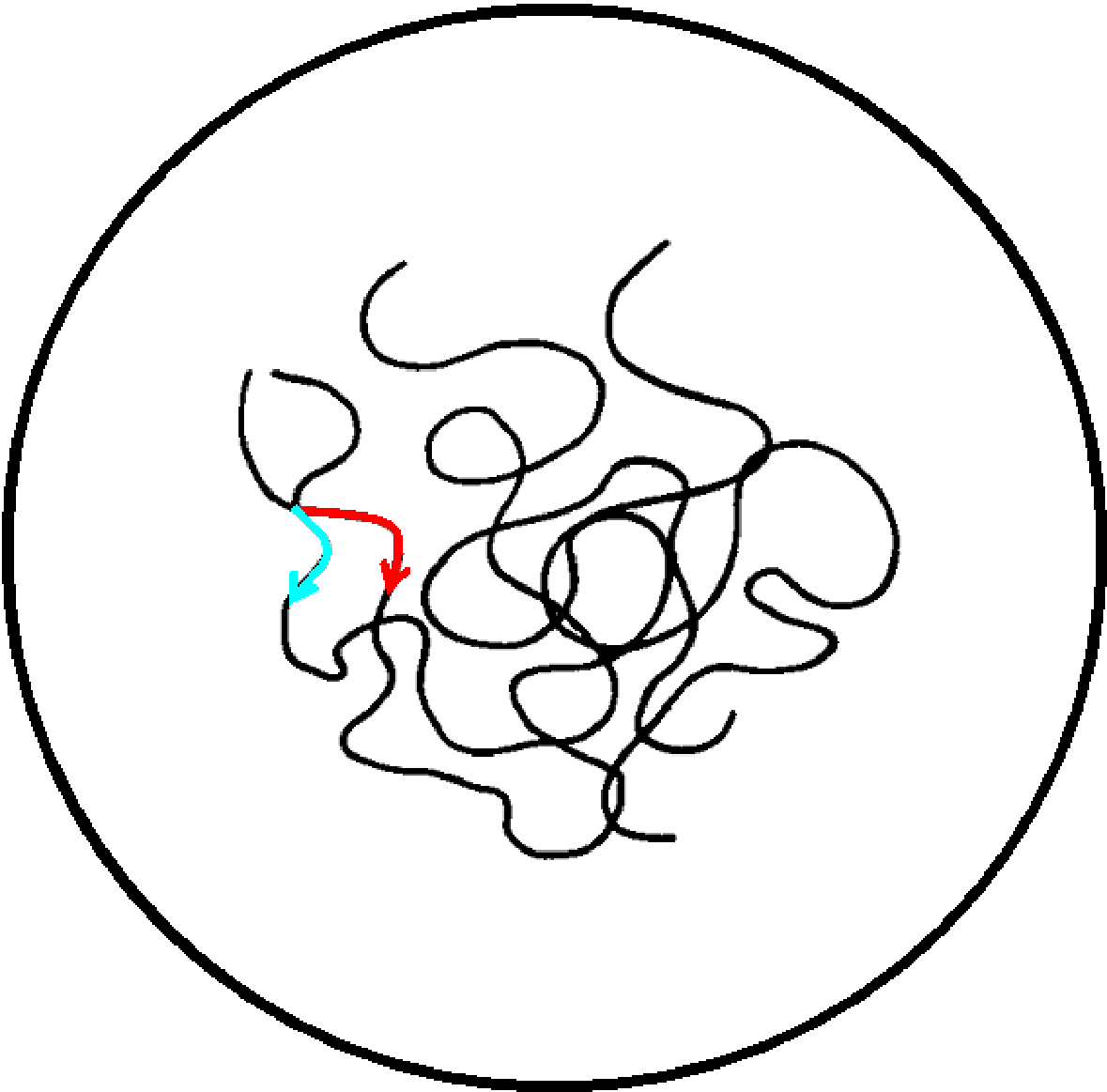}
\caption{A perturbation of bits of string in the same spacetime region but
on different strings (or on well-separated parts of the same string) will  diverge by propagating on the one-dimensional strings.}\label{fig:smallVsHankelPPS}
\end{center}
\end{figure*}

One might wonder about the (prospective) physical  mechanism  that underlies rapid scrambling
in a bound state of strings. Any perturbation of the bound state away from
its equilibrium configuration must propagate along a trajectory
that follows the path of one of the strings.
Nevertheless, spacetime  points nearby one another can live on
different strings (or on distinct portions of the same long string)
and, therefore, nearby modes will tend to propagate along completely different
trajectories following completely different directions. In this way,
a perturbation of one small region of the string state can quickly spread
out to a much larger region because of the divergent paths of the resulting waves. This effect is depicted in Fig~\ref{fig:smallVsHankelPPS}.

Despite being very far apart on a single long string, string bits can be very close in spacetime and therefore can interact. In other words, the string interactions are effectively non-local from the string point of view and a disturbance can be expected  to  delocalize on the string in a short time scale. But is this ``delocalization
time''  the  same as the  minimally short time scale that is saturated by  a perturbed BH?

In general, the scrambling time can be associated with the time scale for the onset of fully  chaotic behavior  in a perturbed dynamical system, and it  is determined  (in part)  by the largest positive Lyapunov exponent $\lambda_L$. (See, {\em e.g.}, \cite{MaldShenk,SwingHayd,JoePetal} for recent  discussions in the context of BHs.) The  exponent $\lambda_L$ appears in an exponentially growing factor  that  is  associated with the degree of delocalization of an initial disturbance. In this case, the  ``delocalization''  is meant to be taking place  in phase space and can be quantified in terms of the  increasing   ``distance''  between once-neighboring trajectories. The maximal amount of delocalization is set by the size of the accessible region of phase space.

In a quantum theory, one of the standard methods  of diagnosing for chaos and then identifying  the Lyapunov exponent is to inspect  an  out-of-time-order four-point function for a suitable choice of unitary operators. With this in mind, we will  apply  an argument from  \cite{SwingHayd} that reveals  how the scrambling time can be deduced for any chaos-prone  quantum theory. The starting point is the following  observation about chaotic motion in phase space: If $q(t;q_0,p_0)$ and $p(t;q_0,p_0)$ are a canonically conjugate pair describing trajectories in  phase space ($q_0$ and $p_0$ are their initial values), it can be expected that
\be
\frac{\partial q(t)}{\partial q_0}\;\sim\; e^{\lambda_L\; t}\;.
\label{chaos}
\ee

Now suppose that
$\;U(t)=e^{i\alpha q(t)}\;$ and $\;V(t)=e^{i\beta p(t)}\;$, where
$\alpha$ and $\beta$ are dimensional constants.  Insofar as  $\alpha$, $\beta$, and $t$ are
suitably small, then Eq.~(\ref{chaos}) plus some  basic quantum mechanics   leads to
\be
\langle U^{\dagger}(t)V^{\dagger}(0)U(t)V(0)\rangle \;\sim\;
\exp\left[-i\alpha\beta\hbar\; \exp\left(\lambda_L t\right)\right]\;,
\ee
where we have temporarily  restored Planck's constant to emphasize that the product $\alpha\beta$ has  units of inverse action. A large phase in an out-of-time-order four-point function can be viewed as a signal of chaotic behavior. Here, the phase becomes order unity
when
\be
t\;\sim \; -\frac{1}{\lambda_L}\ln{(\alpha\beta\hbar)}\;.
\ee

The  scrambling time $t_{scr}$ can  be identified as the time for which  a disturbance is delocalized throughout
phase space. Therefore, $t_{scr}$ can be estimated  by identifying the measure of action  $(\alpha\beta)^{-1}$ with the size of
the phase space in question.  Our model has a  phase space of size  $N\hbar$ since the number of interacting degrees of freedom is $N$ and they all interact with each other in a democratic way.
Consequently,
\be
\label{scrambtime}
t_{scr}\;\sim \; \frac{1}{\lambda_L}\ln{N}\;.
\ee

What is left is to determine the Lyapunov exponent $\lambda_L$. To do this, we will adopt a commonly used method from statistical mechanics which relates the value of the free energy to the Lyapunov exponent. In the standard treatment, one calculates the transfer matrix  $T$ (see, {\em e.g.}, Section~10 of \cite{KogutWilson}, as well as \cite{creutz}) and determines its largest eigenvalue $\lambda_+$. The Lyapunov exponent $\lambda_L$ can  then be identified with the  logarithm  of this largest eigenvalue,
 $\lambda_L=\ln \lambda_+$  (see, {\em e.g.}, \cite{transfer}). The argument uses two different representations of the partition function $Z$,
\be
Z\;=\; {\rm Tr}\; T_1^N
\ee
and
\be
Z\;=\; {\rm Tr}\; e^{-\beta F}\;,
\label{partition}
\ee
where $\beta$ is the inverse temperature and $T_1$ is the transfer matrix for a single degree of freedom.
The transfer matrix acts in Euclidean time over an interval of length $\beta$.

Let us denote the number of degrees of freedom (more generally, the  entropy) by $N$. Then, for large $N$,
\be
Z\;\sim\; \lambda_+^N \;=\; e^{\beta\lambda_L N}\;,
\label{transfer}
\ee
where $\;\lambda_+=e^{\beta \lambda_L}\;$.
Comparing Eqs.~(\ref{partition}) and (\ref{transfer}), we see that the Lyapunov exponent can then be expressed in terms of the free energy per degree of freedom $F/N$,
\be
\beta \lambda_{L}\;=\; -\frac{\beta F}{N} \;.
\ee

Let us recall that the
free energy of the polymer  can be expressed at equilibrium as
\be
\label{freeE}
-\frac{F}{T_{Hag}N}\;=\;\epsilon - \frac{1}{2} \frac{g^2N}{V}\;=\;
\frac{1}{2}\epsilon\;
\ee
and that $\epsilon=1/R_S$. It can therefore be concluded that
\be
\lambda_{L}\;=\; \frac{1}{2}\epsilon\;=\frac{1}{2 R_S}\;.
\ee

Using the standard definition of  the Hawking temperature
$\;T_{Haw}= 1/ (4 \pi R_S)\;$ (which also follows from the free energy in Eq.~(\ref{freeE})), we have
\be
\label{lyapexp}
\lambda_{L}\;=\; {2\pi}{T_{Haw}}\;.
\ee
And so the  Lyapunov exponent can be identified with the inverse of the Hawking thermal wavelength.

We would like to emphasize that our derivation of the Lyapunov exponent follows from its standard definitions for dynamical systems and does not rely on the ratio  of redshifts between different observers. Indeed, gravity never enters into our calculation.

A positive value for $\ln{\lambda_{+}}$ means that the maximal eigenvalue of the transfer matrix is larger than unity and serves as a standard indicator of chaotic behavior. Further, for a strongly interacting system at a finite temperature, the Lyapunov exponent   should be close to but  bounded from above by the inverse of the  thermal wavelength \cite{MaldShenk}. For our model,  this bound appears to be saturated, just as expected for a BH.

One  can  now be conclude, using Eqs.~(\ref{lyapexp}) and~(\ref{scrambtime}), that the  scrambling time for our model is indeed the anticipated expression \cite{HayPres,Suss,Suss2},
\be
t_{scr}\;=  \; \frac{1}{2\pi T_H}\ln{S_{BH}}\;.
\ee

An alternative method for estimating the Lyapunov exponent  goes as follows: The inverse of the Lyapunov exponent is associated with the time scale for a system to equilibrate or thermalize. We can  estimate this time scale by considering a  perturbation away from equilibrium of the distribution of strings for some length of string $\ell_0$. Let us then recall  the Boltzmann-like equation~(\ref{Boltz2}) and consider
$\;n(\ell_0)\rightarrow n(\ell_0)+\delta n(\ell_0)\;$.
It can be shown that such a perturbation dissipates in time  according
to  \cite{LT}
\be
\frac{\delta n(\ell_0,t)}{\delta n(\ell_0,0)} \;=\;  e^{-a\ell_0\frac{g^2}{V}
\left(N+\frac{\ell_0}{2}\right)t}\;\simeq\; e^{-a\ell_0 \frac{g^2N}{V}t}\;,
\label{diss}
\ee
where $a$ is some order-unity numerical factor in string units.

The longest equilibration time for the string state should be
determined by adding the shortest possible string, which means setting  $\;\ell_0=1\;$
in the exponent of  Eq.~(\ref{diss}).
It follows that
\be
\lambda_L \;=\; a\frac{g^2N}{V}\;=\; a\epsilon\;=\;  \frac{a}{R_S}\;.
\ee
Causality requires that $\;a \le 1\;$, and so
\be
\lambda_L \;\lesssim \; \frac{1}{R_S}\;.
\ee
The previous discussion indicates that $\;a=1/2\;$.

\subsection{Viscosity to entropy ratio}

The discussion in the previous subsection regarding dissipation will enable us to estimate  the shear viscosity $\eta$ for the stringy  fluid within the interior. The viscosity can, in general, be identified with the ratio of the energy density
$\;\rho=\epsilon^2/g^2\;$ and the time scale for  dissipation $\lambda_L$. So that, in our case,
\be
\eta\;\simeq\;\frac{\rho}{\lambda_L}\;=\; \frac{2}{\epsilon}\frac{\epsilon^2}{g^2}\;,
\ee
where the second equality follows from Eqs.~(\ref{bhm}) and~(\ref{lyapexp}). Using Eq.~(\ref{consol}), we then  have
\be
\eta\;\simeq\;\frac{2\epsilon}{g^2} \;=\; 2\frac{N}{V}\;=\;2 s\;.
\ee

Our conclusion is that  the shear viscosity  is the same order as the entropy density, $\;\eta/s \simeq 1\;$.  This is true not only at the horizon, as expected \cite{membrane,PSS,KSS}, but throughout the entirety of the interior region!

It is amusing to consider the following method for fixing $\eta$ with
precision.  For certain physical situations, the shear viscosity
of a fluid can be determined  by
an expression of what is basically  the form (see, {\em e.g.}, \cite{sv})
\be
\eta\;=\; 2\frac{T}{\gamma^2}\lambda_L\; s\;,
\ee
where $\gamma$ is the shear rate which has dimensions of velocity divided by length.

The maximum possible  shear rate  is when the velocity is that of light and the length in question is that of the system. In our case, the maximal shear rate would then be given by the inverse radius
\be
\label{gammamax}
\gamma_{max}=1/R_S\;.
\ee
It could be expected, based on the previous discussion, that  the stringy fluid does indeed saturate the maximal value of the shear rate. Then, substituting for $\lambda_L$ from Eq.~(\ref{lyapexp}) and setting $\;T=T_{Haw}\;$, we find that
\be
\eta \;=\; \frac{1}{4\pi} s\;,
\ee
which precisely saturates the celebrated  bound of Kovtun, Son and Starinets \cite{KSS}. It follows that the maximal value of $\lambda_L$ corresponds to the minimal value of $\eta$, just as one would anticipate.

This  uniform saturation of what is believed to be an absolute lower bound on the viscosity-to-entropy ratio is a novel prediction of our model. It also further supports the notion that the interior of the BH is a place where semiclassical physics
breaks down, as the saturation of this lower bound is supposed to be a signature of strongly coupled field theories \cite{MaldShenk}.

\section{Conclusion}

We have revisited our recent proposal that the BH interior is a bound state of long, highly excited, closed strings or, equivalently,
a string-theory version  of  a  collapsed polymer. After reviewing some of the model's salient features, we
introduced  several new consistency checks for this paradigm. We showed that
the position  of the outer surface fluctuates quantum mechanically with the expected relative variance
of $1/S_{BH}$. We also demonstrated that the process of  Hawking radiation ---  with  its anticipated rate of emission  and energy per emitted particle ---   can be attributed
to the quantum effect of small strings  escaping
from the main body of the polymer .
In other words, the BH  horizon emerges as a natural feature in the
polymer framework.

In addition, we have applied standard methods from statistical mechanics and chaos theory and found values for the Lyapunov exponent, scrambling time and  viscosity-to-entropy
ratio that saturate their respective bounds and, therefore, agree with
the standard expected results for BHs. A novel feature of our model is
that the viscosity-to-entropy ratio saturates its  minimally allowed value
throughout the interior and not just at the horizon.
One of our next objectives \cite{collision} is to find a novel prediction that could potentially
be verified  by the (anticipated)  data coming from  gravitational-wave detectors  \cite{LIGO}.

\section*{Acknowledgments}

We would like to thank Yshai Avishai, Doron Cohen, Guy Gur-Ari, Masanori Hanada and Sunny Itzhaki for valuable discussions.
The research of RB is supported by the Israel Science Foundation grant no. 1294/16. The research of AJMM received support from an NRF Incentive Funding Grant 85353, an NRF Competitive Programme Grant 93595 and Rhodes Research Discretionary Grants. AJMM thanks Ben Gurion University for their  hospitality during his visit.

\end{document}